# THE ADVANCED PHOTON SOURCE INJECTOR TEST STAND CONTROL SYSTEM

J.F. Maclean and N.D. Arnold, Advanced Photon Source, Argonne National Laboratory, Argonne, IL USA

Abstract

The Advanced Photon Source (APS) primary and backup injectors consist of two thermionic-cathode rf guns. These guns are being upgraded to provide improved performance, to improve ease of maintenance, and to reduce downtime required for repair or replacement of a failed injector. As part of the process, an injector test stand is being prepared. This stand is effectively independent of the APS linac and will allow for complete characterization and validation of an injector prior to its installation into the APS linac.

A modular control system for the test stand has been developed using standard APS control solutions with EPICS to deliver a flexible and comprehensive control system. The modularity of the system will allow both future expansion of test stand functionality and evaluation of new control techniques and solutions.

## 1 INTRODUCTION

The Advanced Photon Source (APS) is a third generation light source that provides high brightness x-ray beams to a user community. The main purpose of the APS injector test stand (ITS) is to test and characterize injectors destined for use in the APS. It is anticipated that the injectors tested will be of varying designs. These requirements dictate that the control system must be easily adaptable for reconfiguration of the test stand.

The first task of the test stand is to test and characterize replacements for the APS main injectors. Since the provision of replacement injectors had some urgency it was necessary to find a way to implement control quickly whilst allowing for expansion and reconfiguration at a later date. It was therefore decided to use standard APS control components where possible but not to preclude new solutions if they were warranted on technical, cost, or other grounds. Once a functioning system was completed, it was anticipated that the test stand would provide a good opportunity to test control solutions that are new to the APS. It would allow testing in an operational environment with the benefit of not impacting the operation of the APS.

## 2 TEST STAND

The test stand is located in its own shielded room, which is approximately 5.5 m by 3 m in size. The beamline is mounted on an optical bench of approximately 3 m by 1.2 m. Items requiring control include many of the elements found in the main APS linac:
- Injector – the gun under test
- Magnets – correctors, quadrupoles, and dipoles
- Vacuum equipment – pumps and valves
- Beam scraper
- Current monitors – transformers and Faraday cups
- Video cameras
- Galvanometers

The beamline is described in [1], and the main user control screen reproduced in Figure 1 gives an indication of the layout of the line. Figure 2 shows the main components of the control system.

## 3 CONTROL SYSTEM

### 3.1 General

The injector test stand control system uses EPICS [2], which is used for the control system of the APS. One input/output controller (IOC) is used to control the ITS. This is a Motorola MVME162 CPU mounted in a 6U VME chassis. The chassis is physically located in a room adjacent to the test stand. The IOC is connected to the APS control local area network (LAN) via a dual redundant Ethernet link. Mounted in a 19-inch rack with the IOC is other ITS control hardware. This includes:
- Video camera driver chassis
- Stepper motor drive chassis
- Allen-Bradley 1771 chassis
- Tektronix TDS7404 oscilloscope

### 3.2 Links to the APS Control System

Although it is to a large extent independent from the APS control system, the ITS does depend upon it for a number of facilities.
- Control LAN and infrastructure. The ITS IOC is connected to the main APS control LAN and servers. This allows control, monitoring, and development from any workstation on the subnet.
- Timing. A signal is sent to the ITS IOC from the APS timing system ten microseconds prior to an rf pulse.
- Video. The video signal from the ITS is sent to the central APS video multiplexor for distribution.

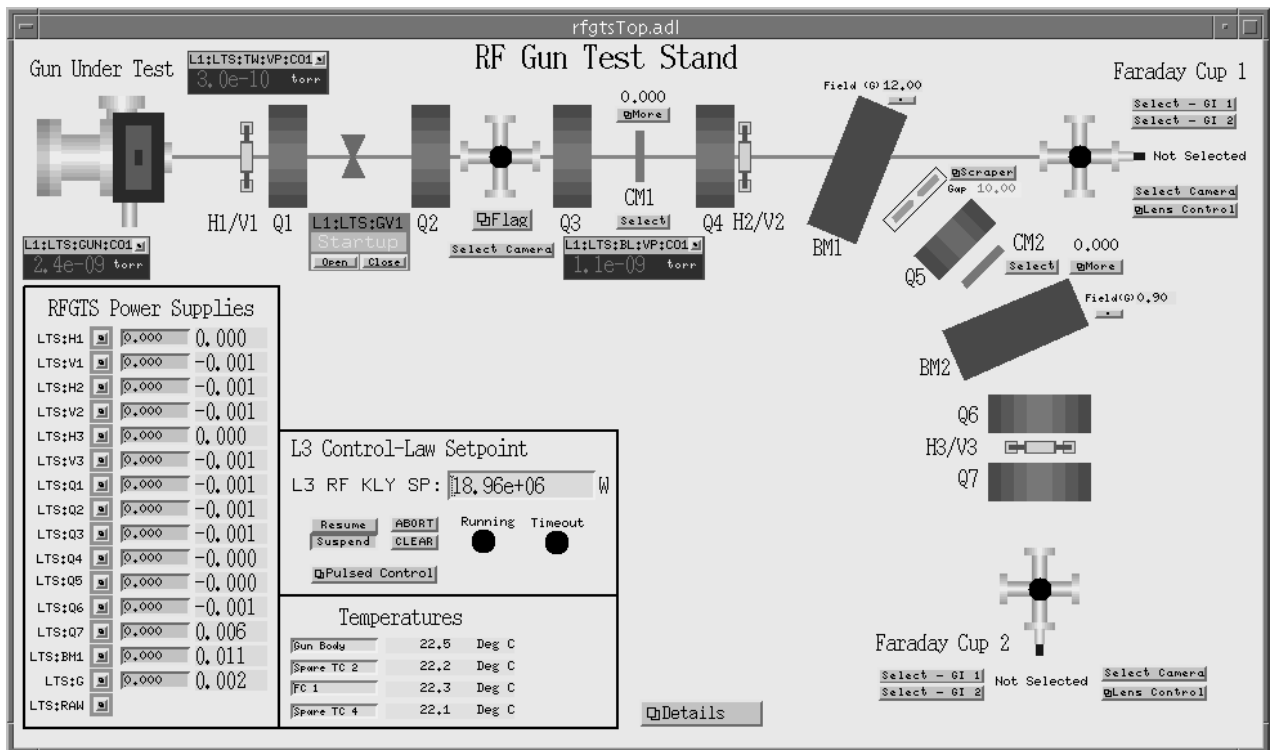
Figure 1. Main ITS control screen, showing main components.

- Interlock. Vacuum or gun cooling faults will cause the rf to trip off.

### 3.3 Input and Output

Input and output (IO) are done via VME cards in the IOC chassis. EPICS support already existed for all IO cards used, and all but one card had previously been used at the APS.

The one VME device new to the APS, a Berkley Nucleonics BN950 digital delay generator, provides a good example of the ITS being used as a test bed for introducing new equipment. In this case, a message to the EPICS "tech-talk" mailing list quickly revealed that EPICS device support had already been written at the Berlin Electron Synchrotron (BESSY). This is an example of a benefit from the EPICS collaboration.

The main individual elements of the ITS and the interface methods used to control them are:
- Magnet power supplies. These are controlled by a standard APS power supply control unit (PSCU) that communicates with the IOC over a Bitbus fiber.
- Vacuum pumps. These are controlled by a standard APS interface unit connected to the IOC by Bitbus fiber.
- Beam scraper. Two scraper blades are driven by stepper motors controlled by an Oregon Micro Systems OMS58 VME card. Positional feedback is provided by linear potentiometers read by a 16-bit analog-to-digital converter industry pack module.
- Current monitors. Four current monitors are multiplexed into two APS gated integrator VME cards. Timing signals are provided via the BN950 delay generator. Spare multiplexor channels are available for future expansion.
- Galvanometers. These are GPIB devices interfaced to the IOC by APS GPIB/Bitbus converters.

### 3.4 Databases

Much of an EPICS-based control system is typically built with databases. The ITS is no exception. Each controlled hardware component has a database associated with it. Since most of the hardware components were of a type that had previously been used in the APS linac, databases already existed for them. We were thus able to reuse software in the form of these databases. Many required little more than the renaming of record instances via macro substitution and the changing of some physical parameters.

This brought the usual benefits associated with software reuse in terms of reduced development time, reduced debugging time, and improved reliability of the finished product.

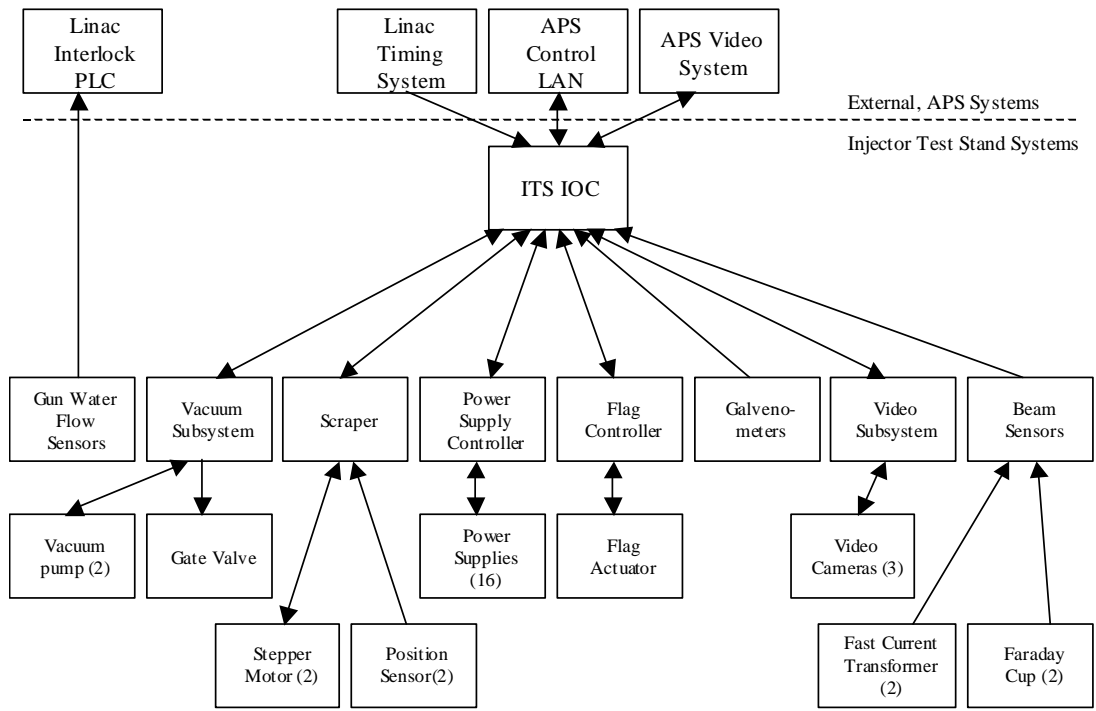

Figure 2: Block diagram showing major ITS control system components.

## *3.5 Displays*

The engineering control screens were produced using MEDM, which is the APS standard user interface tool. This was another area that permitted some software reuse. Control screens for some elements, such as the scraper, power supplies, and galvanometers, were reused from screens originally created for use in other areas of the APS. Configuration of process variable names is achieved by passing parameters when launching the screens.

## 4 RESULTS AND FUTURE WORK

An effort was made to integrate the control system to the hardware as soon as the hardware was available. The success of this effort was shown by the fact that beam was obtained from the first injector tested on the test stand less than two hours after rf energy was applied to it. This included the time required for "rf conditioning" of the injector. In subsequent testing of a second injector, beam was obtained in less than one hour after applying rf power.

APS physicists were very pleased with the ease with which the control system was brought into use. This can largely be attributed to the reuse of software modules and the reliability of EPICS.

Besides tests of new injectors we intend to use the ITS for tests of new control equipment, including a replacement for the standard APS power supply control unit.

## 5 ACKNOWLDGEMENTS

Many people contributed to the ITS control system. The authors would particularly like to acknowledge the hard work and assistance of the following people: William Berg, Richard Koldenhoven, John Lewellen, Josh Stein, and Jim Stevens.

This work is supported by the U.S. Department of Energy, Office of Basic Energy Science, under Contract No. W-31-109-ENG-38.